\documentclass[MSNbibl,nameyear,dvips]{arxstspdf}
\usepackage{flushend}
\usepackage{stfloats}
\volume{29}
\issue{1}
\pubyear{2014}
\firstpage{36}
\lastpage{41}
\doi{10.1214/13-STS424} 



\begin{document}
\begin{frontmatter}

\title{From Science to Management: Using Bayesian Networks to Learn
about~\textit{Lyngbya}}
\runtitle{From Science to Management: Using BNs to Learn about \textit{Lyngbya}}

\begin{aug}
\author[a]{\fnms{Sandra} \snm{Johnson}\corref{}\ead[label=e1]{sandra.johnson@qut.edu.au}},
\author[b]{\fnms{Eva} \snm{Abal}\ead[label=e3]{e.abal@uq.edu.au}},
\author[c]{\fnms{Kathleen} \snm{Ahern}\ead[label=e4]{kattyahern@gmail.com}}
\and
\author[d]{\fnms{Grant} \snm{Hamilton}\ead[label=e2]{g.hamilton@qut.edu.au}}
\runauthor{Johnson, Abal, Ahern and Hamilton}

\affiliation{Queensland University of Technology, University of Queensland,
Department of Environment and Heritage Protection, and Queensland
University of Technology}

\address[a]{Sandra Johnson is Ph.D., School of Mathematical Sciences, Queensland University of Technology,
GPO Box 2434, Brisbane, QLD 4001, Australia \printead{e1}.}
\address[b]{Eva Abal is Associate Professor, Office of the Deputy
Vice Chancellor-Research, University of Queensland,
Brisbane, QLD 4072 Australia \printead{e3}.}
\address[c]{Kathleen Ahern is Ph.D., Department of Environment
and Heritage Protection, 400 George St, Brisbane, QLD 4000, Australia \printead{e4}.}
\address[d]{Grant Hamilton is Ph.D., School of Earth, Environmental and Biological
Sciences, Queensland University of Technology, GPO Box 2434, Brisbane,
QLD 4001, Australia \printead{e2}.}

\end{aug}

%
\begin{abstract}
Toxic blooms of \textit{Lyngbya majuscula} occur in coastal areas
worldwide and have major ecological, health and economic consequences.
The exact causes and combinations of factors which lead to these blooms
are not clearly understood. \textit{Lyngbya} experts and stakeholders
are a particularly diverse group, including ecologists, scientists,
state and local government representatives, community organisations,
catchment industry groups and local fishermen. An integrated Bayesian
network approach was developed to better understand and model this
complex environmental problem, identify knowledge gaps, prioritise
future research and evaluate management options.
\end{abstract}

%

\end{frontmatter}

\section{Introduction}

One of the most common marine pests in waterways around the world is
algae. Harmful algal blooms occur across the world and have a wide
range of detrimental impacts (\cite{hamilton2009}). For example, they
can replace or degrade other algal species that act as fish breeding
grounds, poison fish and mammal marine life through the production of
toxins (\cite{arthur2006}; \cite{arthur2008}), adversely affect coastal
economies through reduced tourism and fishing (\cite{watkinson2005}),
and affect human health through dermatitis
(\cite{osborne2001}; \cite{osborne2007}), neural disorders and
contamination of
other seafood such as shellfish (\cite{pittman2005}). One of the most
common forms of harmful algae is cyanobacteria, or blue-green algae,
and one of the most common species of cyanobacteria in tropical and
subtropic coastal areas worldwide is \textit{Lyngbya majuscula}
(\cite{dennisonetal99}; \cite{arquitt2004}). \textit{Lyngbya}, also
known as
mermaid's hair, stinging limu or fireweed, appears to be increasing in
both frequency and extent (\cite{dennison1999}; \cite{albert2005}).
These blooms
are due to a complex system of biological and environmental factors,
exacerbated by human activities (\cite{watkinson2005}). Thus, while
there is a wealth of scientific and social literature on different
aspects of the \textit{Lyngbya} problem, for example, the role that
nutrients play in the initiation and extent of \textit{Lyngbya} blooms,
or the effect of industry practices in the catchment on the nutrients
available for \textit{Lyngbya} growth, effective management of \textit{Lyngbya}
requires a ``whole-of-system'' approach that comprehensively integrates
the different scientific factors with the available management options
(\cite{johnsonLyngbya}). There is also a need to understand the
different factors that trigger the initiation of a bloom versus the
sustained growth of the cyanobacteria bloom.

Bayesian models are natural vehicles for describing complex systems
such as these (\cite{johnson12ibn}). Key attributes of Bayesian models
in this context include flexibility of the model structure, the ability
to incorporate diverse sources of information through priors and the
provision of probabilistic estimates that take appropriate account of
uncertainty in the system (\cite{mccann2006}; \cite{jensen2007};
\cite{hamilton2009}).
A Bayesian network (BN) is a graphical Bayesian model that uses
conditional probabilities to encode the strength of the dependencies
between any two variables (\cite{pearl1985}). Causal and evidential
inferential reasoning may be performed by the BN, depending on the
nature of the dependencies (\cite{pearl1985}). BNs are increasingly used
to model complex systems (\cite{bromley2005}). Variables in the model
are represented by nodes, and links between variables are represented
by directed arrows. Each node is then ascribed a probability
distribution conditional on its parent nodes. The information used to
develop these distributions can be obtained from a variety of sources,
including data relevant to the system, related experiments or
observations, literature and expert judgement (\cite{mccann2006};
\cite{jensen2007}). A common practice is to discretise the variables
into a
set of states, resulting in a series of conditional probability tables;
hence, under the assumptions of directional separation (d-separation,
so that the nodes are conditionally independent) and the Markov
property (so that the probability distribution of a node depends only
on its parents), the target response node is quantified as the product
of the cascade of conditional probability tables in the network
(\cite{uusitalo2007}). The quantified model can then be used to identify
influential factors, perform scenario assessments, identify
configurations of node states that lead to optimal response outcomes
and so on. BNs can be expanded into object-oriented and dynamic
networks (\cite{jensen2007}; \cite{johnsonLyngbya}); they can include extensions
such as decision, cost and utility nodes (\cite{jensen2007}); and they
can be linked to other BNs to create systems of systems models.

In this paper we describe an integrated Bayesian network (IBN) approach
developed by our research team to address the problem of
\textit{Lyngbya} blooms in Deception Bay, Queensland, Australia. With
its proximity to Brisbane, Australia's third largest city, Deception
Bay is a popular tourist destination in the Moreton Bay region. The
many waterways feeding from intensive and rural agricultural activities
into the bay and its use for commercial and recreational fishing put
pressure on the marine environment and compound the issues resulting
from a \textit{Lyngbya majuscula} bloom (\cite{dennison1999}). Our
project was undertaken as part of the \textit{Lyngbya} Management
Strategy funded by the local and Queensland Government's Healthy
Waterways Program. The project team comprised a \textit{Lyngbya}
science working group and a \textit{Lyngbya} management working group,
representing diverse scientific disciplines, industry groups,
government agencies and community organisations. The IBN is now a
living part of the Healthy Waterways Program and has been expanded
beyond Moreton Bay.

\section{An Integrated Bayesian Network for \textit{Lyngbya}}

The IBN approach that we developed involved a ``science model'' linked to
a ``management model''. The components of the IBN are detailed below.

\subsection{The Science Model}

The science BN [depicted in Supplemental Figure~1
(\cite{johnson2013supp})] comprised the target node, ``Bloom
Initiation'', and 22 other nodes which were identified by the
\textit{Lyngbya} science working group as potentially playing a key
role in the initiation of a \textit{Lyngbya} bloom
(\cite{johnson2013supp}). It was transformed into an
object-oriented BN with subnetworks describing water (comprising nodes
for past and present rain, groundwater and runoff), sea water (tide,
turbidity and bottom current climate), air (wind and wind speed), light
(surface light, light quality, quantity and climate) and nutrients
(dissolved concentrations of iron, nitrogen, phosphorus and organics,
particulates, sediments nutrient climate, point sources and available
nutrient pool) (\cite{johnsonLyngbya}). The nodes of the science model
were quantified using a range of information sources and models,
including process and simulation models, Bayesian hindcasting models,
expert elicitation, published and grey literature, and data obtained
from monitoring sites, industry records, research projects and
government agencies (\cite{johnsonLyngbya}).

\subsection{Science Model Extensions, Alternatives and Sub-Models}
The science object-oriented BN model was further extended to
incorporate temporal trends through a dynamic Bayesian network
comprising five time slices, one for each of the summer months November
to March (\cite{johnson2009}). Lag effects of rainwater and groundwater
runoff were incorporated in the object-oriented BN, allowing
information and influence from one month to flow through to the next
(\cite{johnson2009}).

Additional BNs were also constructed to more fully evaluate the \textit
{Lyngbya} problem. These included separate BNs to model \textit
{Lyngbya} biomass, duration and decay (as opposed to
initiation), and a BN to focus on the critical two month summer period
in which most \textit{Lyngbya} initiations occur (as opposed to annual
averages of rainfall and temperature used in the
original model).

A variety of other statistical models were used to quantify some of the
nodes of the BN.
For example, random forest models were created to predict benthic
photosynthetically active radiation (\cite{kehoe2012}) and Bayesian
regression models were developed
using data obtained from the monitoring stations in the catchment (\cite
{hamilton2009}). The latter data set comprised \textit{Lyngbya}
occurrences for each month during January 2000 to May 2007, a total of
77 observations, and
monthly averages of minimum and maximum air temperature (as proxies for
water temperature), solar exposure and amount of sky not covered by
cloud (as proxies for light), and total
rainfall (as a proxy for nutrients available in the water column),
measured over the same period.
A Bayesian probit time series regression model was developed to predict
the monthly probability of bloom based on a total of 17 covariates,
comprising five main effects, five first-order
autoregressive terms and seven selected interactions. Covariate
selection was performed using a Bayesian reversible jump Markov chain
Monte Carlo algorithm and Bayesian model averaging was used to obtain a
final predictive model. Eight of the 890 models identified by the
algorithm accounted for over 75\% of the posterior model probability,
and the model comprising a single term, average monthly minimum
temperature, accounted for almost 50\%.

\subsection{The Management Model}

The aim of the management network [Supplemental Figure 2 (\cite
{johnson2013supp})] was to facilitate evaluation of options available
to government agencies, communities and industry groups that could
potentially influence the delivery of nutrients to Deception Bay.
\mbox{Nutrient} point sources, such as industries (e.g., aquaculture, poultry)
and council facilities (e.g., waste water treatment plants),
and diffuse sources, such as landuse (e.g., grazing land, forestry) and
urban activities (e.g., stormwater), were geographically located in the
catchment. Each of these sources was then quantified with respect to
the probability of high or low emissions of different types of
nutrients under current, planned and best practice scenarios. While not
a Bayesian network in the
sense of propagating these probabilities, the network structure was a
valuable vehicle for collating and displaying this information.

A GIS-based nutrient hazard map for the catchment was then developed
for each unit of land in the catchment, based on the nutrient emissions
of the sources, the soil pH and soil type at each source location, and
distance of the sources to the nearest waterway (\cite{pointon2008}).
This included a nutrient risk rating which was interpreted as the
perceived risk that there will be ``enough'' of that nutrient to cause an
increase in growth, extent and duration of a \textit{Lyngbya} bloom.

\subsection{Creating and Using the Integrated Bayesian Network}

The science BN and the management network described above were
integrated via a water catchment simulation model that was developed as
part of the \textit{Lyngbya} project. The IBN was
conceived as a series of steps, whereby a management intervention is
proposed, and the management model is used to inform about the expected
nutrient discharge into the Deception Bay catchment. The catchment
model simulates the movement of these nutrients to the \textit{Lyngbya}
site in the Bay, and the science network then integrates this nutrient
information with the other factors in the BN to determine the
probability of bloom initiation.

We briefly discuss here three ways in which the IBN was interrogated to
learn about \textit{Lyngbya majuscula} bloom initiation in Deception Bay.
First, the science BN provided an overall probability of \textit
{Lyngbya} bloom initiation based on the BN structure and its inputs.
For example, in a typical year, as defined by the \textit{Lyngbya}
management working group, the probability of a bloom was predicted to
be $0.28$. Based on the dynamic network, this probability was much
higher in the months of November and December and fell slightly in March.

Second, the IBN informed about important factors affecting this
probability. For example, based on the science network, the seven most
influential factors were available nutrient pool (dissolved), bottom
current climate, dissolved iron, dissolved phosphorus, light and
temperature. Based on both networks,
the comparative impact of different management land uses on the
probability of a bloom could be computed: these probabilities were
lowest for waste water treatment plant ($0.23$) and grazing ($0.27$),
and highest for waste disposal ($0.63$), aquaculture ($0.63$) and
poultry ($0.62$).

Third, the IBN facilitated the evaluation of scenarios, for example,
about the impacts of management options such as upgrading nominated
point sources from current to best practice (e.g., eliminating
potassium output
from sewage treatment plants), climate events (e.g., a~severe storm)
and conditions most or least favourable for bloom initiation. For
example, under optimal light climate and high temperature
conditions, a storm event increased the probability of bloom initiation
from $0.28$ to $0.42$ and initiation was certain if the available
nutrient pool (dissolved) was enough.
As another example, changing the management land use from natural
vegetation to agriculture throughout the catchment area (based on the
management network) results in an increase of 8.8\% in available
nutrients compared with baseline levels (based on the GIS hazard map),
which in turn results in a substantial increase in the probability of a
\textit{Lyngbya} bloom initiation to $0.62$ (based on the science BN).
Note that the effect of this land use change is diluted by the fact
that the proportion of the catchment designated as natural vegetation
is only 18.24\%.

Investigation of the BN also revealed unexpected results that required
discussion and reflection by the science and management teams.
For example, the model supported early suggestions that iron was a key
nutrient in \textit{Lyngbya} bloom initiation (\cite{watkinson2005}),
which motivated additional research into this important issue (\cite
{ahern2008}).
As another example, land runoff and point sources contributed
approximately equally to the probability of bloom initiation under the
developed science model, provoking questions about the relative
effects of population pressure and industrial growth in the catchment.
Alternatively, it suggests that the information available to quantify
these nodes is somewhat uncertain.
In fact, it is a methodological challenge to accurately model the
nutrient load into Deception Bay from land runoff (\cite{kehoe2012})
and more accurate models are currently under development.

\section{Why Bayesian?}

By their nature, a complex system is challenging to model---using
traditional statistical approaches. This is illustrated well in the
\textit{Lyngbya} case study described here, which is characterised by
multiple interacting factors drawn from science and management,
piecemeal knowledge and diverse information sources (\cite{kehoe2012}).
Furthermore, Bayesian models are able to capture the uncertainty in the
data and parameter estimates which is generally agreed to be lacking in
many ecological modelling paradigms (\cite{hamilton2009}). More
specifically, Bayesian networks (BNs) are capable of diagnostic,
predictive and inter-causal (or ``explaining away'') reasoning
(\cite{jensen2007}; \cite{johnson12ibn}), which was particularly
relevant for
the \textit{Lyngbya} problem described here.

There are several alternatives to the IBN approach that could be
considered for modelling the \textit{Lyngbya} problem.
\citet{janssens2006} proposed a decision tree approach, but this was
less able to represent the many interactions between the factors in the
system. Other methods include stochastic petri nets which are able to
model concurrent systems (\cite{angeli2007}), but require the modeller
to have advanced statistical knowledge and were unlikely to engage the
diverse group of \textit{Lyngbya} stakeholders. Process-based
modelling, which is commonplace in ecology, requires substantial data
for calibration and validation of the models, which is very time
consuming and resource hungry and may take several years
(\cite{kehoe2012}). In contrast, a BN allows us to assimilate current
knowledge and modelling effort without having to wait until ``perfect''
and ``sufficient'' data are available. This is particularly important
when dealing with a major environmental hazard such as toxic algal
blooms. None of the alternative approaches had the unique combination
of qualities of BNs which integrated the different sources of
information, represented the dependencies and uncertainty in the
information, guided future data collection and research, and engaged a
diverse group of stakeholders.

The IBN described in this paper is the most comprehensive local systems
model of \textit{Lyngbya} that has been developed to date. There are
many other examples of the use of BNs to solve ``big'' problems. We have
employed them to investigate infection control in medicine, airport
and train delays, wayfinding, import risk assessment
(\cite{mengersen11}), peak electricity demand and sustainability of the
dairy industry in Australia. Furthermore, there are other conceptual
and methodological approaches to \mbox{constructing} BNs; examples include
decision making in business (\cite{baesens2004}) and protein networks in
biology (\cite{jansen2003}).

Finally, BNs are just one tool in the kit of statistical methods that
should be considered for solving these types of problems and that can
be considered as complements to other approaches in order to reveal the
full picture of a complex system.

\section*{Acknowledgement}

The authors gratefully acknowledge Kerrie Meng\-ersen for her significant
contribution to this manu\-script.

\begin{supplement}
\stitle{Supplementary Figures\\}
\slink[doi,text={10.1214/13-\break STS424SUPP}]{10.1214/13-STS424SUPP} 
\sdatatype{.pdf}
\sfilename{sts424\_supp.pdf}
\sdescription{Diagrams for the \textit{Lyngbya} management network and
the \textit{Lyngbya} science Bayesian network are included in the
supplemental article to this paper.}
\end{supplement}


\end{document}